\documentclass[12pt]{article}

\usepackage{amsmath}
\usepackage{amssymb}

\newcommand{\om}{\omega}
\newcommand{\prt}{\partial}

\begin{document}

\title{Theory of rapid (nonadiabatic) rotation of nonspherical nuclei
\footnote{Zh. Eksp. Teor. Fiz. {\bf 73}, 785--801 (1977) [Sov. Phys. JETP
{\bf 46,} No.~3, 411-420 (1977)]}}

\author{V.G. Nosov$^{\dagger}$ and A.M. Kamchatnov$^{\ddagger}$\\
$^{\dagger}${\small\it Russian Research Center Kurchatov Institute, pl. Kurchatova 1,
Moscow, 123182 Russia}\\
$^{\ddagger}${\small\it Institute of Spectroscopy, Russian Academy of Sciences,
Troitsk, Moscow Region, 142190 Russia}
}

\maketitle

\begin{abstract}
On the basis of the concept of the growing role of nonadiabatic effects of the
non-conservation of the quantum number $K,$ a theory has been developed of the
phenomenon which has been given the name of backbending. Above the transition point,
for $J\geq J_c$, all the values $-J\leq K\leq J$ are equally probable. An
investigation is made of the singularities possessed by the ordering parameter
(proportional to the spectroscopic quadrupole moment of a nonspherical nucleus),
the rotational angular velocity and the moment of inertia of a nucleus at the
Curie point.  Formulas have been derived for the intensity of quadrupole radiation
in the more symmetric $n$-phase $J> J_c$. By analyzing the experimental values of
the moments of inertia belonging to the  $n$-phase, the radius of the mass
distribution in the nucleus was determined. It agrees with the radius of the
proton distribution derived from data on the scattering of electrons by nuclei.
On the basis of the simplest form of the singularity of the parametric derivative
of the Hamiltonian of the system a general theory of zero-temperature second-order
phase transitions is developed in the Appendix.
\end{abstract}

\section{Introduction. Estimate of the critical value $J_c$ of the rotational
quantum number. Rotational density matrix}

In recent decades there perhaps has not been in nuclear physics an event more
outstanding than the discovery made at the beginning of the 1950's by A.~Bohr and
B.~Mottelson of rotational levels in nonspherical nuclei.   Having confirmed as a
subsidiary result the well-known assertion of quantum mechanics concerning the
inability of a perfectly spherical body to undergo purely mechanical rotation
(cf., for example, Refs.~[1,2]), this discovery became the starting point for a
detailed study of the so-called rotational bands in the energy spectra of
nonspherical nuclei.   The initial simplest theoretical treatment is expressed
by the widely known formula
\begin{equation}\label{1}
    E(J)\cong\frac{\hbar^2}{2I'}J(J+1)
\end{equation}
for the energy levels.   Here $J$ is the rotational quantum number, i.e., the
total angular momentum of the whole system (nuclear spin); $I'$ is the ``adiabatic"
moment of inertia which refers to the neighborhood of the origin of the band and
the value of which is determined from experiment.   Avoiding as far as possible
complications which do not have particular significance as to principle, we shall
in future as a rule have in mind the case of a rotational band based on the ground
state of an even-even nucleus.   Then in fact the allowable values are $J=0,2,4,\ldots$

Progress of the experiment upward along the band stimulated an ever more critical
attitude to formula (1).   One of the methods of generalizing it consists of the
following:  we adopt the point of view which appears to be natural that the conserved
quantity $J$ is the only significant physical (thermodynamic) characteristic of which
the energy $E$ is a function within the band under consideration in the case of not
too light a nucleus.
Any scalar constructed from the components of the vector $\mathbf{J}$ will be reduced
to the combination $J(J+1)$. Therefore
\begin{equation}\label{2}
    E_m(J)=\sum_{s=1}^\infty C_s[J(J+1)]^s
\end{equation}
(the reason for indexing the rotational energy of the nucleus by $m$ will become
clear from subsequent discussion).    From a practical point of view doubt immediately
arises whether neglecting terms of the series (2) after a finite their
number would lead to a situation that is sufficiently stable from a purely
computational point of view.   However with time an even deeper difficulty involving
matters of principle became apparent.   About six years ago due to the application of
new experimental methodology penetration began into the domain of still greater values
of $J$ [3-6]. This led to the discovery of a very characteristic phenomenon which was
called backbending \footnote{This is due to the external appearance of curves that
illustrate the position of rotational levels of the band and constructed in terms
of somewhat artificially chosen coordinates.   In the present work we shall not
make use of such coordinates.}.
When a noticeable effect was observed it consisted of the following:
the principal characteristics
\begin{equation}\label{3}
    \frac{dE}{dJ}=\hbar\Omega,\qquad \frac{d^2E}{dJ^2}=\frac{d\Omega}{dJ}=\frac{\hbar^2}I
\end{equation}
($\Omega$ is the rotational angular velocity, $I$ is the moment of inertia) of the
rotational motion of a nucleus undergo a sharp change over a very narrow region of the
rotational band.   The angular velocity of rotation $\Omega$ here falls off precipitously
in spite of an increase in the total angular momentum $J$.   It is also very characteristic
that in some cases immediately beyond the transition point the rotational spectrum turns
out to be close to an equidistant one.   For example, quite typical in this regard is the
rotational band based on the ground state of the nuclide W$^{170}$.   The segment
$J=10-14$ in this case contains only two intervals.   But over this segment the angular
velocity $\hbar\Omega$ (measured in energy units) manages to decrease by $\sim 100$ keV---%
by approximately 30\%.   The sector $J= 12-16$ which is minimally displaced in the
positive direction also contains two energy intervals.   But the second of them exceeds
in the first one by only 3\%, and the increase in angular velocity of rotation here is
just as insignificant (cf., formula (3)).

At first glance it might appear that the discovery of the phenomenon of backbending
compromises the very basis for arguing in favor of relations of the type (1) or (2).
However it appears to us to be not quite logical and excessively hasty to cast doubt
on the determining role of the independent (and, with macroscopic accuracy, essentially
unique) thermodynamical variable $J$. In particular, the arguments leading to formula
(2) by no means exclude the possibility that this series would have a finite radius of
convergence.   In accordance with the spirit of experimental data the point $J=J_c$
of the intersection of the boundary of the circle of convergence with the real axis
could then have an important physical meaning.   If nuclear states situated on opposite
sides of this point differ by some qualitative property of a symmetry type, then for
$J > J_c$ the energy $E(J)$ must be a different function which does not coincide with
the analytic continuation of expression (2).   Due to the presence of a singularity at
$J = J_c$ such an analytic continuation could, strictly speaking, turn out to be not
single-valued and even complex.

In order to discuss the specific physical nature of the phenomenon we turn to the
properties of the component $\mathbf{J}\cdot\mathbf{n}=K$ ($\mathbf{n}$ is a unit
vector along the axis of the non-spherical nucleus) of the total angular momentum of
the nucleus along its symmetry axis.   Although this is not a matter of principle
and does not in any manner affect exact calculations, in order to make the subsequent
discussion more easily visualizable one can, if one so wishes, call upon a crudely
classical vector model; one can set $J_z=(J_z)_{max}=J$ and then $K\cong J\cos\theta$.
The quantum number $K$ is not conserved since the operator $\mathbf{J}\cdot\mathbf{n}$
does not commute with the Hamiltonian of the system.   However for $J\ll J_c$ near
the origin of the rotational band the commutator is small.   It is well known that
this corresponds to adiabatic slowness of rotation.   As a result we obtain
\begin{equation}\label{4}
    K\cong K_0=\mathrm{const},\qquad w_K\cong \delta_{KK_0},\qquad J\ll J_c,
\end{equation}
where $w_K$ is the probability of having the corresponding value of $K$ in the quantum
state under consideration of the nucleus as a whole.   Here formula (1) is valid.
In practice we most frequently have to deal with the simplest case $K_0=0$ for even-even
nuclei.

But as one moves along the rotational band the admixture of components with
$K\neq K_0$ increases more and more rapidly.   A rough estimate making use of
perturbation theory leads to a relation of the type
$$
w_K\propto [2|K|/(J-|K|)]^{J-|K|}e^{J-|K|}
$$
to express the law of growth of a given component near the threshold $J= |K|$
where it first appears.   High above the threshold we obtain the expression
$w_K(J)\propto J^{2|K|}$, which remains sufficiently steep and also favors the appearance
of large $K\sim J$.   As long as perturbation theory is applicable the initiating
component $K = K_0$ appears to serve, in agreement with unitarity, as a source of growth
for the others.   However, when the latter cease being relatively small, then all the
component become qualitatively on the same footing in the process of their, so to say,
interaction with each other.   The result will be a kind of equilibrium, an
equidistribution with respect to $K$ ($w_K = 1/(2J+ 1) = \mathrm{const}$), which will be
attained, say, at the point $J=J_c$.   Of course, here one should not expect any
appreciably sharp change in the state of the system as such, for example a discontinuity
in its energy.   Physically the important feature of such a new situation is contained in
a different attribute---in the change in the symmetry of the rotational state of the nucleus.
It can be easily understood that an equidistribution with respect to $K$ corresponds to
isotropy, to an equal probability of all spatial directions of the vector $\mathbf{n}$ (we recall
that we are dealing with an individual specific quantum state of the nucleus as a whole:
$J$ and $J_z$ are fixed quantities!).   At the same time in the Hamiltonian of the system
as a whole one also should not be able to discern any favored directions for the vector
$\mathbf{n}$ in empty space.   It would be naive to think that this correspondence will be
violated in the course of a further increase in $J$ when physically there are even fewer grounds
for some values of the quantum number $K$ to become favored compared to others.   Having
acquired stability the more symmetric isotropic phase will not lose it for $J>J_c$.

The attainment of isotropy with respect to the directions of $\mathbf{n}$ can also be
interpreted from a somewhat different point of view.   In a sufficiently strong rotational
``field" the mechanical angular momenta of the individual quasiparticles aline parallel to
the vector $\mathbf{J}$ and cease to aline along the vector $\mathbf{n}$.   Therefore the
latter turns out to be ``free," i.e., it is in fact distributed isotropically (for
$J \geq J_c$).  This constitutes the difference from the old adiabatic region $J \ll J_c$
where the internal state of the nucleus is determined, roughly speaking, by its deformation
and, in turn, determines the approximately conserved value $K = \mathbf{J}\mathbf{n}\simeq K_0$.
At the transition point itself the number of aligned quasiparticle is of order unity, i.e.,
\begin{equation}\label{5}
    J_c\sim l\sim k_fR\gg 1
\end{equation}
($k_f$ is the limiting momentum of the Fermi distribution; $R$ is the nuclear radius).
This estimate does not contradict experimental data (cf., also Sec.~5).

Thus, $J=J_c$ is the point of a phase transition of the second kind---the zero-temperature
Curie point.   In analogy with the terminology adopted in Ref.~7 we shall refer to the states
of a nonspherical nucleus with a completely isotropic distribution of its axis in space as
the $n$-phase, and we shall call the region $J<J_c$ of ``spontaneous symmetry violation" the
$m$-phase. We then have
\begin{equation}\label{6}
    w_K=\frac1{2J+1},\quad \overline{\cos^2\theta}=\frac13,\quad
    \overline{K^2}=\frac1{2J+1}\sum_{K=-J}^J K^2=\frac{J(J+1)}3,
    \quad J\geq J_c
\end{equation}
($\theta$ and $\varphi$ are the spherical angles defining the direction of the vector
$\mathbf{n}$ with respect to fixed axes).   We emphasize that these properties of the
$\mathbf{n}$-phase presuppose a most essential indeterminancy in the value of the polar
angle $\theta$ which does not in any way diminish with increasing $J$.   In the $m$-phase
the relations (6) are violated as a result of the ``ordering" of the rotational state.
Having in mind, primarily, the special case $K_0 = 0$ and taking into account the fact
that in the adiabatic region $J \ll J_c$ (cf., formulas (1) and (4)) a change in the
quantum number $J$ has little effect on the quantity $\overline{K^2}$, we shall, in
order to be specific, set
\begin{equation}\label{7}
    \overline{K^2}=\sum_{k=-J}^J w_KK^2<\frac{J(J+1)}3,\qquad J<J_c.
\end{equation}

It can be easily verified that for a state which is a superposition of components with
different values of $K$ it is not possible to construct any purely rotational wave
function which depends only on $\mathbf{n}$.   For $J \gtrsim J_c$ the quantum mechanical
description of the rotation of a nonspherical nucleus is achieved with the aid of the
density matrix $\rho(\mathbf{n}, \mathbf{n'})$.   However, due to the conservation of the
$z$-component of the total angular momentum the ``azimuthal" wave function
\begin{equation}\label{8}
    \phi_{rot}=(2\pi)^{-1/2}e^{iM\varphi},\quad M=J_z.
\end{equation}
nevertheless exists.   In the quasiclassical case $J_z=(J_z)_{max}=J\gg 1$ it corresponds
to the so-called regular precession of an ordinary symmetric top (cf., for example, Ref.~8).
As regards the polar angle $\theta$, sufficiently far above the Curie point the density
matrix which depends upon it must have the following properties:  $\rho(\theta,\theta')=0$
for $\theta\neq\theta'$, while the angular dependence of the diagonal element of
$\rho(\theta, \theta')$ corresponds to isotropy in three-dimensional space.   However the
formula appears more attractive if one utilizes a representation which is based not on
angles, but on the more customary quantum numbers $J$, $M$ and $K$.   Then we have
\begin{equation}\label{9}
    \rho_{J_0J_z}(J,M,K;J',M',K')=\frac1{2J_0+1}\,\delta_{JJ_0}\delta_{J'J_0}\delta_{MJ_z}
    \delta_{M'J_z}\delta_{KK'}
\end{equation}
in a state of angular momentum $J_0$ and its component $J_z$.   For diagonal elements
the region of applicability of this relation is wider and encompasses the whole $n$-phase
including the Curie point.   But for $J\to J_c+0$, apparently, non-diagonal (with respect
to $K$) elements of the density matrix which reflect the correlations between different
$K$ also become significant.   It is even more difficult to form judgements concerning the
specific form of the rotational density matrix in the $m$-phase $J <J_c$.   For $J = J_c$
it has some kind of a singularity which hinders the formal continuation of the corresponding
expression into the ``foreign" region $J > J_c$. The physically important functional of
the rotational density matrix---the so-called order parameter---has a radical singularity
as $J\to J_c-0$ (cf., next section).

\section{The average (``spectroscopic'') quadrupole moment as an order parameter.
Discontinuity in the rotational angular velocity of the nucleus.}

If we keep (6) and (7) in mind, the following definition suggests itself
\begin{equation}\label{10}
    \eta=1-\frac{3\overline{K^2}}{J(J+1)}\cong 1-3\overline{\cos^2\theta}
\end{equation}
for the order parameter which characterizes deviation of the $m$-phase from isotropy
(the approximate expression involving the angle refers to the quasiclassical case
$J_z=J\gg1$ and is stated here only for ease of visualization).   In the important
special case $K_0 = 0$ it in fact varies from unity to zero at the Curie point.
Naturally, for a given $J$ and $\eta$ quite different nuclear states are in principle
conceivable---both rotational states, and ones differing microscopically.   However,
we shall assume that for $J < J_c$ one can utilize the concept of the energy of that
one of these states which is ``most nearly in equilibrium" and which we shall denote
by $E(J, \eta)$.

Intending to analyze the situation for sufficiently small $J_c-J$ in the neighborhood
of the Curie point we expand the general expression for the energy in powers of $\eta$
and limit ourselves to three terms:
\begin{equation}\label{11}
    E(J,\eta)=E_n(J)-A(J)\eta+\tfrac12D(J)\eta^2.
\end{equation}
The function $E_n(J)$ in fact refers to the unordered $n$-phase.   The inequalities
\begin{equation}\label{12}
    A(J)>0,\qquad D(J)\cong D(J_c)\equiv D>0
\end{equation}
express the advantages of positive $\eta$ and the stability of the energy minimum
with respect to this quantity. Substitution into (11) of the true value of $\eta$
determined from the equilibrium condition
\begin{equation}\label{13}
    \frac{\prt E}{\prt\eta}=0
\end{equation}
transforms $E(J, \eta)$ into the energy $E_m(J)$ of the rotational levels of the $m$-phase
\begin{equation}\label{14}
    \eta=\frac{F(J)}D,\qquad E_n(J)-E_m(J)=\frac{[A(J)]^2}{2D}.
\end{equation}

Figure 1, which shows the curve  for the energies of the phases, requires some explanation.
At the ``end-point" $J=J_c$ the function $E_m(J)$ has a certain singularity. However, the
question is:  Which of the derivatives of the function will be first affected by this, i.e.
which of them will become infinite at this point?  According to (3) this cannot happen in
the case of the first derivative:   there are no slightest theoretical or experimental
bases for supposing that the rotational velocity of nuclei becomes infinite.   Moreover,
as may be seen from
formula (3), the second derivative also does not become infinite.   A zero value of the
moment of inertia agrees neither with the fact that the nucleus is non-spherical, nor with
the experimentally observed tendency for moments of inertia in the $m$-phase to vary
(cf. Sec.~5).   Thus, although the dotted line extrapolation of the $E_m(J)$ curve is
somewhat arbitrary, at least the first two derivatives have meaning along it.   As a result
of this assumption illustrated by Fig.~1 that at $J=J_c$ a simple point of intersection
\begin{equation}\label{15}
    E_n-E_m\propto J_c-J
\end{equation}
of the curves under investigation occurs is not contradictory and agrees well with the
nature of the experimental data.

Comparison of (15) with (14) yields
\begin{equation}\label{16}
    A(J)=a(J_c-J)^{1/2},\qquad a>0.
\end{equation}
Consequently
\begin{equation}\label{17}
    \eta=a(J_c-J)^{1/2}/D
\end{equation}
and
\begin{equation}\label{18}
    E_n-E_m=\frac12\frac{a^2}D(J_c-J).
\end{equation}
Differentiating relation (18) we obtain a discontinuity in the rotational angular
velocity of the nucleus at the point of ``backbending"
\begin{equation}\label{19}
    \Delta(\hbar\Omega)=a^2/D.
\end{equation}
Here $\Delta(\hbar\Omega)\equiv\hbar\Omega_{mc}-\hbar\Omega_{nc}$.   As far as
one can judge according to experimental data, the thermodynamic inequality
\begin{equation}\label{20}
    \Omega_{mc}>\Omega_{nc}
\end{equation}
is never violated.

It is well known [1,2] that for the value averaged over the state $J_z=J$ of the
$z$-component of the quadrupole moment (the so-called spectroscopic quadrupole
moment) of a nonspherical nucleus the following formula is valid
\begin{equation}\label{21}
    Q=\frac{3\overline{K^2}-J(J+1)}{(J+1)(2J+3)}\,Q_0.
\end{equation}
Here, $Q_0$ is the component of the ``collective'' macroscopic quadrupole moment
along the nuclear axis wholly determined by its axially-symmetric deformation.
Taking (10) into account we have
\begin{equation}\label{22}
    Q=-\frac{J}{2J+3}\,Q_0\eta.
\end{equation}
Now comparing this with (17) we verify that both the quadrupole moment $Q$ and
also the order parameter vanish at the Curie point according to $(J_c-J)^{1/2}$.
In the $n$-phase $J\geq J_c$ there can no longer be any collective (spectroscopic)
quadrupole moment because of its isotropy.

\section{Position of the rotational levels in the $n$-phase $J>J_c$. Moment of
inertia.}

First of all we touch in a couple of words upon the problem of the stability of
either of the phases. The corresponding condition
\begin{equation}\label{23}
    \frac{d^2E}{dJ^2}>0
\end{equation}
has a fairly natural appearance
\footnote{It is not difficult to verify that negative moments of inertia correspond to
absolute instability both thermodynamic and purely mechanical.   The latter in
the present case has the following meaning:  free rotation of a body corresponds,
as is well known, to thermodynamic equilibrium and is not accompanied by friction
(cf., for example, Refs.~9 and 10).   If the regular precession with respect to
the angle $\phi$ is treated in accordance with the laws of classical mechanics
(cf. the Introduction, formula (8) and the text referring to it), then for negative
moments of inertia the principle of least action is violated.}.
The symmetric $n$-phase becomes unrealizable for $J < J_c$ because of the absolute
instability of the state in which the mechanical angular momenta of the quasiparticles
aligning  along the vector $\mathbf{J}$ ignore the direction of the vector $n$ determined
by the deformation (cf., the in a certain sense opposite nature of the states which are
in fact realized in the adiabatic region $J\ll J_c$; cf., also the Introduction).
Therefore we have
\begin{equation}\label{24}
    \left.\frac{d^2E_n}{dJ^2}\right|_{J=J_c}=0
\end{equation}
and
\begin{equation}\label{25}
    \frac{d^2E_n}{dJ^2}\propto J-J_c
\end{equation}
near the Curie point.   Substitution into the second of
formulas (3) yields
\begin{equation}\label{26}
    I\cong\frac{j}{J-J_c},\qquad J-J_c\ll \frac{j}{I_0}.
\end{equation}
Here $j$ is a certain constant coefficient;
\begin{equation}\label{27}
    I_0=\frac25MR^2
\end{equation}
is the solid body value of the moment of inertia; $M = m_nA$ is the nuclear mass;
$m_n$ is the nucleon mass;
\begin{equation}\label{28}
    R=r_0A^{1/3}
\end{equation}
is the nuclear radius.   We arrive at an important conclusion that the moment
of inertia of a nonspherical nucleus becomes infinite in accordance with the
law (26) if one approaches the Curie point from above.   This is what explains
the observed approximate uniform spacing of levels in corresponding segments of
the rotational spectra (cf., the Introduction).   In the opposite limiting case
of large $J - J_c$, for a sufficiently clearly pronounced lining-up of mechanical
angular momenta of the quasiparticles along the direction of the vector
$\mathbf{J}$ the moment of inertia of the Fermi system becomes the same as that
for a solid body:
\begin{equation}\label{29}
    I\cong I_0,\qquad J-J_c\gg \frac{j}{I_0}.
\end{equation}

If the right-hand sides of (26) and (29) are added, then an interpolation formula
will be obtained for the moment of inertia of the $n$-phase, which satisfies both
limiting cases.   By itself it, of course, is devoid of any to some extent deep
physical foundations.   Moreover, in the case of a literal acceptance of this
interpolation a situation would arise of a monotonic approach to the solid body
limit $I_0$ from above, while experimentally, apparently, a minimum was observed
in the variation of the moment of inertia (cf., Sec.~5).   However, in a purely
technical respect the interpolation formula is a convenient route for obtaining
correct limiting expressions for $\Omega_n(J)$ and $E_n(J)$ by means of integrating
it twice.   Retaining everywhere in addition to the terms containing constants of
integration, a single term of the corresponding expansion we obtain the following
formulas
\begin{equation}\label{30}
    \begin{split}
    \hbar\Omega&=\hbar\Omega_{nc}+\frac{\hbar^2}{2j}(J-J_c)^2,\\
    E&=E_0+\hbar\Omega_{nc}(J-J_c)+\frac{\hbar^2}{6j}(J-J_c)^3
    \end{split}
\end{equation}
for $J-J_c\ll j/I_0$ and
\begin{equation}\label{31}
    \begin{split}
    \hbar\Omega&=\hbar\Omega_{nc}+\frac{\hbar^2}{I_0}(J-J_c),\\
    E&=E_0+\hbar\Omega_{nc}(J-J_c)+\frac{\hbar^2}{2I_0}(J-J_c)^2
    \end{split}
\end{equation}
for $J-J_c\gg j/I_0$.

It is desirable to keep in mind that, generally speaking, the last formula predicts
the position of the rotational levels with good relative but not absolute accuracy.
We note that for nuclides for which $j/I_0\sim 1$ the domain of applicability of
formulas (31) is broadened encompassing practically the whole $n$-phase.

\section{Electric quadrupole radiation in the $n$-phase}

The basic property of the $n$-phase can be formulated as the absence of an observable
(average) quadrupole moment;
\begin{equation}\label{32}
    \langle Q_{ik}\rangle =0.
\end{equation}
This does not mean that its matrix elements non-diagonal with respect to the levels
of the band also vanish.   They are responsible for the quadrupole radiation the
probability of which we can write in the form
\begin{equation}\label{33}
    w=\frac{e^2\omega^5}{90\hbar c^5}\langle\langle Q^2\rangle\rangle.
\end{equation}
We have for the sake of brevity symbolically denoted by
$\langle\langle Q^2\rangle\rangle$ (reduced intensity of the transition) the square
of the absolute value of the matrix element between the final state $J-2$ and the
initial state $J$ summed over the final orientations of the nuclear spin. The transition
frequency of interest to us amounts to $\om=2\Omega$.

For the subsequent discussion it is convenient to express this collective quadrupole
moment in terms of the components of the vector $\mathbf{n}$:
\begin{equation}\label{34}
    Q_{ik}=\tfrac12 Q_0(3n_i n_k-\delta_{ik}).
\end{equation}
From this it can be seen that it is immaterial over which state the square of the
tensor is averaged:
\begin{equation}\label{35}
    Q_{ik}^2=\tfrac32 Q_0^2.
\end{equation}
With the aid of the usual rules for matrix multiplication we verify the validity
of the relation
\begin{equation}\label{36}
    \sum_b\left|(Q_{ik})_a^b\right|^2=\tfrac32 Q_0^2
\end{equation}
Here the indices $a$ and $b$ enumerate the individual states of a nonspherical nucleus
irrespectively of to which rotational band they happen to belong.

Above the Curie point the quadrupole radiation can be calculated quasiclassically.
Setting $\phi=\Omega t$ we calculate in accordance with (34) the classically varying
quadrupole moment.   Then in accordance with the isotropy of the $n$-phase it is
averaged uniformly over $\cos\theta$ ($K=J \cos\theta$ in the quasiclassical case).
Finally the square of the absolute value of the spectral component $\om=2\Omega$ is
substituted in formula (33) in place of $\langle\langle Q^2\rangle\rangle$. As a
result of elementary calculations we obtain \footnote{Since the operator for collective
quadrupole moment is diagonal with respect to $K$, the correlations mentioned at
end of the Introduction do not affect the result.}
\begin{equation}\label{37}
\langle\langle Q^2\rangle\rangle=\tfrac14 Q_0^2,\qquad w=\frac{e^2\om^5}{360\hbar c^5}
Q_0^2,\qquad J>J_c.
\end{equation}
Thus, within the bounds of the $n$-phase all the radiative transitions have the same
reduced intensity.   In addition to the transition responsible for the real radiation
there also exists an analogous transition in the opposite direction upwards along
the band.   The sum of the reduced intensities of both transitions is obtained by
doubling the right hand side of the first of formulas (37).   Comparing this with
the sum rule (36) we find that it is saturated to one-third within the band under
consideration, while the remaining two-thirds refer to transitions to other
rotational bands.

\section{Comparison with experiment}

Data on the states of nonspherical even-even nuclei with high spins were taken
primarily from Ref.~11. The variation of the rotational angular velocity
$\hbar\Omega_m$ in the $m$-phase was extrapolated graphically into a relatively
small region near the Curie point (cf., also comments on Fig.~1 in Sec.~2), while
the latter was identified with the position of the energy interval which already
belongs, as far as the available data enable us to judge, to the $n$-phase.
The results of this treatment of data are shown in Table~1.   The two cases marked
by an asterisk refer to the rotational band originating from the $0^+$ level of
the excitation of $\beta$ oscillations (the fully symmetric oscillation of a
nonspherical nucleus).   The quite unsystematic variations in the magnitude of
the increment $\Delta(\hbar\Omega)$ of the rotational velocity from one nucleus
to the next are noteworthy.   The critical angular velocity $\hbar\Omega_{mc}$
behaves in a much more stable manner, but it also is subject to random fluctuations.
It is somewhat more difficult to judge how real on the average is the tendency
to a certain amount of decrease in $\hbar\Omega_{mc}$ with increasing atomic weight.

\begin{table}
\begin{tabular}{|c|c|c|c|c|c|c|c|}
\multicolumn{8}{c}{\bf Table 1}\\
\hline
Nucleus & $J_c$& $\hbar\Omega_{mc}$  &$\Delta(\hbar\Omega_{mc})$ &
Nucleus & $J_c$& $\hbar\Omega_{mc}$  &$\Delta(\hbar\Omega_{mc})$  \\
 & &  MeV & MeV &
 & & MeV & MeV \\
\hline
Ba$^{126}$ & 13 & 530 & 200 & Er$^{162}$ & 15 & 345 & 70 \\
Ce$^{130}$ & 11 & 410 & 175 & Er$^{164}$ & 17 & 370 & 120 \\
Ce$^{132}$ & 13 & 490 & 235 & Er$^{166}$ & $\geq$15 & 310 & $\geq$20 \\
Ce$^{134}$ & 11 & 600 & 365 & Yb$^{164}$ & 15 & 350 & 110 \\
Gd$^{154}$ & $\geq$17 & 340 & $\geq$25 & Yb$^{166}$ & 15 & 340 & 90 \\
Gd$^{154*}$ & 13 & 275 & 75 & Yb$^{168}$ & 15-17 & 330 & $\sim$30 \\
Gd$^{156}$ & $\geq$15 & 310 & $\geq$15 & Yb$^{170}$ & 17 & 365 & 60 \\
Dy$^{154}$ & $\geq$15 & 320 & $\geq$45 & Hf$^{168}$ & 15 & 340 & 110 \\
Dy$^{156}$ & $\geq$17 & 370 & $\geq$40 & Hf$^{170}$ & 15-17 & 300 & $\sim$5 \\
Dy$^{156*}$ & 13 & 275 & 95 & Hf$^{172}$ & $\geq$17 & 360 & $\geq$40 \\
Dy$^{158}$ & 15 & 335 & 45 & W$^{170}$ & 13 & 320 & 105 \\
Dy$^{160}$ & 15 & 310 & $\sim$25 & W$^{176}$ & $\geq$15 & 350 & $\geq$40 \\
Er$^{156}$ & 13 & 380 & 120 & Os$^{182}$ & 15 & 340 & 100 \\
Er$^{158}$ & 15 & 360 & 125 & Os$^{184}$ & $\geq$13 & 385 & $\geq$30 \\
Er$^{160}$ & 15 & 345 & 70 & Os$^{186}$ & $\geq$15 & 460 & $\geq$210 \\
\hline
\end{tabular}
\end{table}

\begin{table}
\begin{tabular}{|c|c|c|c|c|c|c|c|c|}
\multicolumn{9}{c}{\bf Table 2}\\
\hline
 & Ce$^{130}$ & Gd$^{154*}$  & Dy$^{158}$ &
Er$^{162}$ & Yb$^{164}$  & Yb$^{166}$  & W$^{170}$ & Os$^{182}$ \\
\hline
$J_c$ & 11.0 & 12.4 & 14.8 & 15.7 & 13.0 & 15.5 & 13.6 & 11.9 \\
$j/I_0$ & 2.11 & 2.12 & 6.76 & 2.63 & 3.98 & 2.11 & 1.88 & 4.20 \\
\hline
\end{tabular}
\end{table}

In addition to those given in Table~1 there are six more nuclides in which rotational
levels have been discovered up to $J_{max}=14-18$:  W$^{172}$, W$^{174}$, Os$^{176}$,
Os$^{178}$, Os$^{180}$, and Th$^{232}$.   No significant effect was observed. However,
let us turn to the rotational velocities attained.   For the three osmium isotopes
mentioned above we have $\hbar\Omega = 280-315$~keV.   This is lower than the values
characteristic of the end of Table~1 and is close to the lower limit of critical
velocities in the whole table.   For both tungsten isotopes $\hbar\Omega_{max} = 300$~keV.
The thorium nucleus in the experiment attained the value $\hbar\Omega= 200$~keV.
A conclusion suggests itself that in the nuclides ``suspected" of not showing the
effect there was simply a lack of sufficient rotational velocity to equalize the
intensities of the components of the wave function with different values of $K$.
Therefore there is every reason to suppose that with a further increase in $J$ the
situation will become clarified and, as usual, a phase transition will be found.
We note, in order to avoid misunderstanding, that even in the case of a vanishingly
small increment $\Delta(\hbar\Omega)$ (a situation close to this already occurred in
the case of Hf$^{170}$, cf., Table~1) it still will be possible to recognize the
$K$-phase by its moment of inertia.

The study of the variation of the moment of inertia of a nonspherical nucleus in both
phases is of considerable interest.   The most characteristic and complete data in
this respect are those concerning the principal rotational band for W$^{170}$; cf.,
Fig.~2.   The increase in the moment of inertia in the $m$-phase can be qualitatively
understood in the spirit of the Le Chatelier-Brown principle as a ``resistance to an
external action''.  Indeed, the rotational perturbation responsible for the growth
in the components with $K\neq K_0$ is determined not directly by the mechanical angular
momentum $J$, but by the angular velocity $\Omega_m$.   Therefore the system tries to
diminish the derivative $d(\hbar\Omega)/dJ$. \footnote{An analogous interpretation can
also be given to the increase in the moment of inertia of the $n$-phase on the
right wing of its curve (cf., Fig.~2)---only here the system resists the alignment
which begins to acquire a macroscopic scale of the mechanical angular momenta
of the individual quasiparticles along the direction of the vector
$\mathbf{J}$ (cf., the Introduction).
This is what explains the existence of the minimum in the variation of the moment
of inertia of the $n$-phase which will be discussed subsequently.}
On the other side of the Curie point in the $m$-phase, the moment of inertia drops
somewhat below $I_0$, passes through a minimum and then rapidly tends to approach the
solid body asymptote from below.   Analogous non-monotonic behavior in the variation of
the quantity $d^2E_n/dJ^2$ was also observed for the nuclide Yb$^{164}$.
An interesting case was encountered in the $\beta$-vibrational band of dysprosium
(Dy$^{160*}$ according to the notation of Table~1).   From all indications here the
very first two experimental points belonging to the $n$-phase turned out to be situated
on opposite sides of the minimum in the moment of inertia. A final verification of the
hypothesis will have to be postponed until the position of the rotational levels of
the band with $J> 18$ is established.

On the whole the data concerning the $m$-phase are so far relatively scarce. When
both experimental values closest to the Curie point belong to the left wing of the
moment of inertia curve formula (26) yields
\begin{equation}\label{38}
    \frac{\hbar^2}{j}=\frac{(\hbar^2/I)_2-(\hbar^2/I)_1}{J_2-J_1},\quad
    J_c=\frac{J_1(\hbar^2/I)_2-J_2(\hbar^2/I)_1}{(\hbar^2/I)_2-(\hbar^2/I)_1}.
\end{equation}

These relations enable us to obtain the ``extrapolated'' (from the domain of the
$n$-phase) position of the point of phase transition, and also to evaluate $j/I_0$.
For the eight nuclides the results are shown in Table~2.   Basically good agreement is
observed with values of $J_c$ shown in Table~1 determined by a different method.
The greatest discrepancy occurs in the case of osmium.   However, it must be stated
that the value of $J_c = 12$ given in Table~2 falls just in the middle of that segment
of the rotational band where a decrease in the rotational angular velocity occurs.
The region occupied by the phase transition is here anomalously wide and encompasses
three energy intervals.   Possibly this is in some way associated with the closeness
of osmium to the point of phase transition of nonspherical nuclei into spherical ones
[7] (cf., also the striking anomalies at the beginning and at the end of Table~1).
We also note that for specific nuclides the values of the ratio $j/I_0$ vary no less
randomly than the increments $\Delta(\hbar\Omega)$.

The universal character of the asymptotic properties of the $n$-phase far from the
Curie point in principle opens a path to the determination of the radius of the mass
distribution in a nucleus.   Unfortunately at the present time such high spins of
rotational states have not yet been attained for which the relation (29) could be
confidently regarded as a sufficiently strict equality. However one can try to
circumvent this difficulty in the following manner.   On the right wing of the moment
of inertia curve for the $n$-phase, where it is increasing, the corresponding values
are lower than the rigid body values,   This very fact provides us with a lower bound on
the possible values of the nuclear radius.   A less rigorous but still quite plausible
assumption consists of the following:   in cases of the type selected for Table 2 the
experimental values of the moment of inertia of the $n$-phase closest to the Curie
point are higher than the rigid body values.   This provides an upper bound on the radius.
Making use of the fact that data are available concerning eight different nuclides we
narrow down as much as possible the bounds of the inequality:
\begin{equation}\label{39}
    1.08\mathrm{F}<r_0<1.11\mathrm{F}
\end{equation}
(1F$=10^{-13}$ cm). Thus, with reasonable accuracy we have
\begin{equation}\label{40}
    r_0=1.1\mathrm{F}.
\end{equation}
This agrees with the measurements of the radius for the distribution of electrical
charge in the nucleus from electron scattering [12]. In the preceding discussion the
value (40) was utilized everywhere in the calculations of the rigid body moment of
inertia using formulas (27) and (28).

For a comparison of formulas (30) and (31) with experiment one must have in the
$n$-phase a sufficiently large number of experimental points referring to the same
nuclide.   The most favorable case is the one of W$^{170}$.   Since the value of
$j/I_0$ given in Table~2 is not large we may use formula (31).   It is not difficult
to understand the reason for the good agreement over the whole $n$-phase demonstrated
in Table~3.   Being quite well founded under the condition $J-J_c\gg j/I_0$ formula
(31) at the same time has a correct interpolation behavior:   as $J \to J_c$ it gives
an experimental value $\Omega_{nc}$ for the rotational angular velocity of the nucleus
which is known to be correct (cf., also the remark at the end of Sec.~3).   In order
to illustrate the application of formula (30) we in contrast select a nuclide with
the greatest value of $j/I_0$; in accordance with Table~2 this will be Dy$^{158}$.
As Table~3 shows, the theory agrees well with experiment.

We now pass to the question of the electric quadrupole radiation from a
non-adiabatically rapidly rotating nucleus.   In a recently published experimental
paper [13] a study was made of the radiation within the limits of the principal
rotational band of Ce$^{134}$.   Two observed transitions belong to the $n$-phase.
For comparison with previous formulas we reproduce the value of for $K=K_0 = 0$:
\begin{equation}\label{41}
    \langle\langle Q^2\rangle\rangle_{K=0}=\frac9{16}\left(1-\frac1J\right)
\end{equation}
(cf., for example, Ref.~[14]; we neglect the corrections $\sim 1/J^2$).   Dividing
the right hand side of the first of formulas (37) by (41) we obtain the reduced
intensity $F$ measured relative to its purely adiabatic value:
\begin{equation}\label{42}
    F=\frac{\langle\langle Q^2\rangle\rangle}
    {\langle\langle Q^2\rangle\rangle_{K=0}}=\frac49\left(1+\frac1J\right).
\end{equation}

\begin{center}
\begin{table}
\begin{tabular}{|c|c|c|c|c|c|}
\multicolumn{6}{c}{\bf Table 3}\\
\hline
\multicolumn{3}{|c|}{W$^{170}$ Nucleus } &
\multicolumn{3}{|c|}{Dy$^{168}$ Nucleus } \\
\hline
J & $\hbar\Omega$ (keV) experim. & $\hbar\Omega$ (keV) theory &
J & $\hbar\Omega$ (keV) experim. & $\hbar\Omega$ (keV) theory \\
\hline
13 & 216.4 & -- & 15 & 289.0 & -- \\
15 & 223.2 & 240 & 17 & 295.5 & 295.7 \\
17 & 265.0 & 273 & 19 & 312.9 & 313.3 \\
19 & 308.3 & 306 & 21 & 339.0 & 342.0 \\
21 & 342.6 & 339 &    &       &       \\
\hline
\end{tabular}
\end{table}
\end{center}

Setting $J = 12$ we have $F_{theor} = 0.48$.   According to Ref.~13,
$F_{exp} = 0.65 \pm 0.13$ for the transition $12^+\to 10^+$, while the next
transition $14^+\to 12^+$ has the intensity $F_{exp} = 0.64\pm 0.28$.

\begin{center}
\begin{table}
\begin{tabular}{|c|c|c|c|}
\multicolumn{4}{c}{\bf Table 4}\\
\hline
Nucleus & $J_c$ & $\hbar\Omega_{mc}$ (keV) & $\Delta(\hbar\Omega)$ (keV) \\
\hline
Ho$^{157}$ & $\geq$37/2 & 355 & $\geq$100 \\
Ho$^{159}$ & $\geq$37/2 & 340 & $\geq$70 \\
Ho$^{161}$ & $\geq$37/2 & 325 & $\geq$55 \\
Er$^{159}$ & 39/2 & 380 & $\sim$25 \\
Yb$^{165}$ & 35/2 & 305 & 70 \\
Lu$^{167}$ & $\geq$27/2 & 300 & $\geq$15 \\
\hline
\end{tabular}
\end{table}
\end{center}

In conclusion we touch upon the case of odd nuclei. Because it is difficult
to sort out the numerous rotational bands until now there are relatively few
data concerning them.   We know that the principal source of experimental
information concerning the problem of interest to us are the quadrupole $\gamma$
quanta corresponding to transitions with $\Delta J=2$.   Therefore the usual
adiabatic ``rotational band" characterized by certain $K_0$ and parity should be
separated into two appropriate bands and treated as distinct bands.\footnote{The
division into two ``sub-bands" is equally well-founded also from a purely
theoretical point of view.   Indeed, one does not expect any considerable
differences in principle from even-even nuclei in the present case, while
it is desirable to exclude the not entirely clear and non-macroscopic effect of the
``alignment" of the odd nucleon.   It is well known that for $K_0=\tfrac12$  a
similar effect occurs even in the adiabatic region where it is more easily
susceptible of being investigated (cf., for example, Ref.~[1]).}
Results referring to six nuclides and based on the experimental papers [15--17]
are given in Table~4.   They give no cause for doubting the universal character
of the phenomenon.

We express our gratitude to A.I.~Baz', I.I.~Gurevich, I.M.~Pavlichenkov and
K.A.~Ter-Martirosyan for discussions of questions touched upon in this paper
and its results.   The authors are particularly grateful to I.M.~Pavlichenkov
for many stimulating discussions of the nature of experimental data concerning
the phenomenon of backbending.

\setcounter{equation}{0}

\renewcommand{\theequation}{A.\arabic{equation}}

\section*{Appendix}

It appears to be of some interest to analyze certain specific features of
zero-temperature phase transitions of the second kind from a more general
point of view.

Let the energy $E$ of the ground state of a macroscopic body and the state
$\Psi(\xi)$ itself depend on a certain thermodynamic quantity $x$, and the
corresponding condition for stability be represented for the sake of brevity
and simplicity in the form
\begin{equation}\label{a1}
    \frac{d^2E}{dx^2}>0,
\end{equation}
analogous to (23).   We shall assume that for $x<x_c$ a spontaneous breaking
of symmetry occurs which is characterized by the ordering parameter $\eta$.
An analytic continuation of the symmetric $n$-phase into the region $x<x_c$
is not physically realizable because of
\begin{equation}\label{a2}
    \left.\frac{d^2E_n}{dx^2}\right|_{x=x_c}=0.
\end{equation}

Formally this can be only the result of some kind of singularity of the paramedic
derivatives of the Hamiltonian $H(x)$ acting on the variables $\xi$ of the system.
In accordance with
\begin{equation}\label{a3}
    H(x)\Psi_i(\xi;x)=E_i(x)\Psi_i(\xi;x)
\end{equation}
it determines the eigenfunctions and the energy eigenvalues (the Latin subscript
enumerates the stationary states of the system which parametrically depend on $x$).
The simplest imaginable possibility is the following: the singularity in the
derivative $dH/dx$ leads to the ``break" in the curve of $E(x)$ of the type shown
in Fig.~1. Then proceeding in the same manner as in Sec.~2, i.e., actually replacing
$J$ in formulas (11) and (14)--(18) by $x$ we obtain the following results:
\begin{equation}\label{a4}
    \begin{split}
    \eta=a(x_c-x)/D,\\
    E_n-E_m=\frac{a^2}{2D}(x_c-x),\\
    \left.\frac{dE_m}{dx}\right|_{x=x_c}-\left.\frac{dE_n}{dx}\right|_{x=x_c}
    =\frac{a^2}{2D}>0.
    \end{split}
\end{equation}
However, we emphasize that so simple a possibility exists only when the following
important condition is satisfied:  the thermodynamic quantity
$\left.dE/dx\right|_{x=x_0}$ must not reduce to a functional of the wave function
$\Psi(\xi;x_0)$. In the opposite case a discontinuity in the functional can only be
the result of a discontinuous variation of the function itself.   But this would
already be a first-order transition and not a second-order phase transition in
which we are interested for which we always have
\begin{equation}\label{a5}
    \Psi_m(\xi;x_c)=\Psi_n(\xi;x_c).
\end{equation}

We consider the case when the quantity $x$ has the properties of a dynamic variable
(this is possible---cf., for example, Ref.~7---although not necessarily so).
In the presence of a velocity $\dot{x}$ an additional kinetic energy proportional
to its square appears, and the total energy can be expressed in the form
\begin{equation}\label{a6}
    \mathcal{E}=\tfrac12B(x)\dot{x}^2+E(x).
\end{equation}
Here $B(x)$ is a mass coefficient.   Suppose that the system is moving in the
negative direction along the $x$ axis.   At first sight the hypothesis seems
plausible that it, generally speaking, will approach the point $x_c$ with a certain
finite velocity $\dot{x}_n$.   But then in the relation
\begin{equation}\label{a7}
    \dot{\eta}=\gamma\dot{x}
\end{equation}
the condition
\begin{equation}\label{a8}
    \gamma=0,\qquad x\geq x_c
\end{equation}
must hold also for $x = x_c$:  the domain of the more symmetric $n$-phase extends
up to the point of phase transition inclusively [9].   Having all this in mind we
apply the law of conservation of energy directly at the Curie point:
\begin{equation}\label{a9}
    \tfrac12B_m\dot{x}_m^2+E_m=\tfrac12B_n\dot{x}_n^2+E_n,\quad x=x_c.
\end{equation}
In accordance with the second of formulas (A.4) (cf., also (A.5)), we ignore the
``potential energy" $E(x)$. The system cannot pass to the left with a finite
velocity $\dot{x}_m$---differentiation of the first of formulas (A.4) shows that
in violation of condition (A.8) at $x = x_c$ the coefficient $\gamma(x)$ would
change discontinuously from zero to infinity.   Thus, $\dot{x}_m = 0$ within the
framework of the foregoing assumption and, consequently,
\begin{equation}\label{a10}
    B_n(x_c)=0.
\end{equation}
Switching on adiabatically smoothly at $t\to\infty$ the variation $x(t)$ of the
coordinate and solving the time-dependent Schrodinger equation with the Hamiltonian
$H(x(t))$ we obtain the well-known expression
\begin{equation}\label{a11}
    B_n(x)=2\hbar^2\sum_{i\neq n}\frac1{E_i-E_n}\left|\left\langle\Psi_i\Bigg|
    \frac{\prt\Psi_n}{\prt x}\right\rangle\right|^2
\end{equation}
for the mass coefficient (in order to ascribe to the notation a more concrete
character we denote by the index $n$ the ground states $i = 0$ within the domain of
the $n$-phase $x>x_c$).   Since all the denominators are positive substitution
into (A.10) yields
\begin{equation}\label{a12}
    \left\langle\Psi_i\Bigg|
    \frac{\prt\Psi_n}{\prt x}\right\rangle =0,\quad x=x_c.
\end{equation}
Regarding the amplitude $\left\langle\Psi_n|{\prt\Psi_n}/{\prt x}\right\rangle$,
it as a result of normalization turns out to be purely imaginary and can be everywhere
made equal to zero with an appropriate choice of the indefinite phase multiplier
$\exp(i\alpha(x)$ in the wave function of the ground state.   Then all the components
of the derivative $\prt\Psi_n/\prt x$ along the basis vectors of the Hilbert space
turn out to have zero values at the Curie point
\begin{equation}\label{a13}
    \left.\prt\Psi_n/\prt x\right|_{x=x_c}=0.
\end{equation}
Expanding the parametric derivative $\prt\Psi_n/\prt x$ in series and now measuring
the coordinate $x$ from the position of the point of phase transition we represent
the mass coefficient of the $n$-phase in the form
\begin{equation}\label{a14}
    B_n(x)=8bx^2,
\end{equation}
where
\begin{equation}\label{a15}
    b=\frac{\hbar^2}4\sum_{i\neq n}\frac1{E_i-E_n}\left|\left\langle\Psi_i\Bigg|
    \frac{\prt\Psi_n}{\prt x}\right\rangle\right|^2.
\end{equation}
For the solution of equations of classical mechanics it is simplest of all to make
use of their first integral (A.6):
\begin{equation}\label{a16}
    x(t)=\left\{
    \begin{array}{ll}
    (\mathcal{E}/b)^{1/4}(\pm t)^{1/2}\quad & \text{if}\quad x>0\\
    \pm(2\mathcal{E}/B_m)^{1/2}t \quad & \text{if}\quad x<0.
    \end{array}
    \right.
\end{equation}
Here the time is so reckoned that at the instant $t = 0$ the system would be
passing through the point $x = 0$ of the phase transition and the energy
$\mathcal{E}$ is measured from the ``potential energy" $E(0)$.   The upper
sign corresponds to motion in the positive direction of the $x$ axis and the
lower sign to motion in the negative direction.
By $B_m$ we have denoted the limiting value of the mass coefficient of the
$m$-phase as $x \to -0$.   The singularity which the solution of (A.16) in fact
has as $x\to+0$ deprives the proportionality relationship (A.7) of a strict
mathematical interpretation directly at the origin.   By this the paradox is
removed concerning the simultaneous retention both of the law of conservation
of energy and the nature of the phase transition as the system passes through
the Curie point.

Strictly speaking the states that have been considered are not entirely
equilibrium states, since the macroscopic motion is accompanied by inevitable
friction. In the general case in order to achieve complete equilibrium it is
necessary to add to the Hamiltonian a term of the form $-\lambda x$ which does
not affect the nature of the phase transition.   Equilibrium at $x=x_0$
corresponds to $\lambda=\left.dE/dx\right|_{x=x_0}$.   If the zero-temperature
second-order phase transitions have to a large extent a common nature, then the
result (A.13) must not depend on whether the quantity $x$ does in fact have
the properties of an autonomous dynamical variable.   Indeed, it is sufficient
to require complete thermodynamic equilibrium---``rest", in order that the
abstract capability in principle of the system to be ``in motion" would in no
way manifest itself physically.   From this point of view the condition (A.13)
imposed on the wave function of the $n$-phase at the Curie point appears to be
of a sufficiently general nature.

We now return to the dynamic case.   The solution (A.16) was classical.   The
question arises whether the limiting expansion (A.14) for the variation of the
mass coefficient of the $n$-phase might not lose its concrete physical content
when quantum effects are taken into account.   In order to check this, we turn
to the uncertainty relation
\begin{equation}\label{a17}
    \Delta\mathcal{E}\Delta x\sim \hbar \dot{x}
\end{equation}
(cf., for example, Ref.~[1]). According to (A.16) we have
$$
\dot{x}=\pm(\mathcal{E}/b)^{1/2}/2x,
$$
i. e.,
\begin{equation}\label{a18}
    \Delta x\sim\left(\frac{\mathcal{E}}{b}\right)^{1/2}\frac{\hbar}{x}
    \frac1{\Delta\mathcal{E}}\gg\frac{\hbar}{(b\mathcal{E})^{1/2}}\frac1x.
\end{equation}
From this it can be seen that the requirement $\Delta x\ll x$ cannot be satisfied
only for
\begin{equation}\label{a19}
    x\lesssim(\hbar^2/b\mathcal{E})^{1/4}\to 0\quad\text{as}\quad\mathcal{E}\to\infty.
\end{equation}
Thus, there exists a possibility in principle to check by a purely classical method
the validity of the expression (A.14) down to the smallest values of $x$.

In the foregoing friction accompanying motion was not taken into account explicitly.
Generally speaking, the question concerning dissipative processes is quite
complicated and requires a more concrete investigation in each individual case.
However, in the region where the $n$-phase is sufficiently close to the point of
phase transition the situation is simplified.   Indeed, projecting the time-dependent
Schr\"odinger equation onto the base vector system defined by (A.3), we establish
that the connection between the amplitudes of states with different $i$ is realized
exclusively by means of coefficients which in addition to the velocity $\dot{x}$ contain
matrix elements of the form $\left\langle\Psi_k|{\prt\Psi_i}/{\prt x}\right\rangle$.
They must also be responsible for the irreversible quantum jumps from level to level
which are the cause of dissipation.   But, as is clear from (A.13), such matrix elements
vanish at the Curie point and near it friction is suppressed. Therefore results which
concern the limiting behavior of the $n$-phase in the neighborhood of the phase
transition point remain valid.

Finally, we briefly touch upon the relation of the foregoing to the special case $x=J$
of nuclear rotation considered in the present paper.   The complete wave function
of a non-spherical nucleus can be written as a product
\begin{equation}\label{a20}
    \Psi_{J_0J_z}(J,M,\xi)=\delta_{JJ_0}\delta_{MJ_z}\psi_{J_0}(\xi).
\end{equation}
By £$\xi$ we have denoted variables commuting with $J$ and $M$ and with one another
(among their number is contained also the macroscopic quantity
$K=\mathbf{J}\cdot\mathbf{n}$).  The wave functions $\Psi_J(\xi)$ depend on $J$ as
on a parameter.   They obey an equation of the form (A.3) in which the role of the
``Hamiltonian" is played by the block diagonal with respect to $J$ and $M$ of the
complete Hamiltonian for the nucleus (parametrically dependent on $J$), while
rotational energy levels serve as eigenvalues.   Since the quantity
$\hbar\Omega_0=\left.dE/dJ\right|_{J=J_0}$ does not reduce to a functional of
$\Psi_{J_0}(\xi)$ the rotational velocity $\Omega$ can undergo a discontinuity at
the transition point as the energy itself $E(J)$ undergoes a continuous variation
along the band, and this is in fact observed.   The discontinuous diminution of
$\Omega$ signifies in essence a change in the ``coupling scheme" of angular momenta
in the system (cf., Introduction).

\end{document}